\documentclass[12pt]{article}
\usepackage{graphicx}
\usepackage{natbib}
\usepackage{rotating}
\usepackage{amssymb}
\title{  A Turbulence Model \\ for Smoothed Particle Hydrodynamics}
\author{ J. J. Monaghan\\
 \small{School
of Mathematics Sciences}\\ \small { Monash University, Clayton 3800, Australia}\\
\small {email: joe.monaghan@sci.monash.edu.au } }
\date{\small {15 November 2009}}
\begin{document}
\maketitle
\vskip 2cm
\begin{abstract}

The aim of this paper is to devise a turbulence model for the particle method Smoothed Particle Hydrodynamics (SPH) which makes few assumptions, conserves linear and angular momentum, satisfies a discrete version of Kelvin's circulation theorem, and is computationally efficient. These aims are achieved.  Furthermore, the results from the model are in good agreement with the experimental and computational results of Clercx and Heijst for two dimensional turbulence inside a box with no-slip walls.  The model is based on a Lagrangian similar to that used for the Lagrangian averaged Navier Stokes (LANS) turbulence model, but with a different  smoothed velocity.  The smoothed velocity preserves the {\it shape} of the spectrum of the unsmoothed velocity, but reduces the magnitude for short length scales by an amount which depends on a parameter $\epsilon$. We call this the SPH-$\epsilon$ model. The effectiveness of the model is indicated by the fact that the second order velocity correlation function calculated using the smoothed velocity and a coarse resolution, is in good agreement with a calculation using a resolution which is finer by a factor 2,  and therefore requires 8 times as much work to integrate to the same time. 
\end{abstract}

\newpage
\section{Introduction}

The subject of this paper is a turbulence model  for  Smoothed Particle Hydrodynamics (SPH, for a review see Monaghan 2005). The SPH method was developed for astrophysical applications but has been progressively extended to problems involving incompressible fluids by using either a slightly compressible model of the fluid (Monaghan 1994), or by algorithms designed to solve the full incompressible equations (Cummins and Rudman 1999, Hu and Adams 2007).  In this paper we use the slightly compressible model.

 Many of the incompressible flow problems to which SPH has been applied involve the generation of turbulence. A common example is a dam break where the flow of the escaping fluid is laminar until it  hits a downstream wall and forms a breaking wave. Another example is the sloshing of a fluid in an oscillating tank where breaking waves are created each cycle of sloshing.  Standard turbulence models such as the $k-\epsilon$ model have been applied to the SPH simulation of dam break problems (Violeau and Issa 2007) with modest success.  In this paper we make use of the ideas associated with the Lagrangian averaged Navier Stokes equations (LANS) explored by Holm and his colleagues (Holm 1999, 2002, Chen et al. 1999,  Cheskidov et al. 2005, Guerts and Holm 2006),   Mohseni et al. 2003,  and  for extensive references see Graham et al. (2007) and Lunasin et al. (2007, 2008).  The basic idea is to determine a smoothed velocity $\widehat { \bf v}$ by a linear operation on the un-smoothed velocity $\bf {v}$, and then determine the Eulerian equations of motion  from  Lagrange's equations for the SPH particles using a Lagrangian where the kinetic energy per unit mass is $\frac12 \widehat {\bf v} \cdot {\bf v}$. 
 
  The average motion of the fluid is determined by $\widehat{\bf v}$ and in the SPH formulation the particles are moved with this velocity. The resulting acceleration equation contains  extra terms which represent the effective stresses induced by the smoothing.  Once the form of the smoothing is chosen these stresses are determined.  The equations of motion conserve energy, linear and angular momentum (in the absence of rigid boundaries and external forces), and they satisfy a discrete version of the circulation theorem.  
    
    To complete the model we add a viscous dissipation term. We use one of the standard SPH viscosity terms which has been tested for Couette flow (Monaghan 2006), and for spin down in a rotating cylinder (Monaghan 2005, Monaghan and Kajtar 2009). We simulate he boundary conditions associated with fixed or moving bodies  by placing boundary particles on the boundaries.  These boundary particles exert forces on the fluid SPH particles. This technique, which is similar to the Immersed Boundary Method of Peskin (1977, 2002), has been shown to give good results for a wide range of problems (Monaghan and Kajtar 2009). The effects of the turbulence on processes such as  thermal transport  can be estimated using the differences between the smoothed and unsmoothed velocities to determine diffusion coefficients. However, we do not study those processes in this paper.
    
    We apply the SPH-$\epsilon$ equations to turbulence in a two dimensional box with no-slip boundary conditions.  This system has been studied both experimentally and numerically by Clerx et al. (1999, 2000), Massen et al. 2002) and is particularly interesting because it shows that the dynamics with no-slip  or stress-free boundary conditions (relevant to many turbulence problems in nature), differs significantly from that with periodic conditions. This includes the formation of vortices, the spectrum, and changes in the angular momentum. The two dimensional problem also has the advantage that the computational demands are very much less than for a three dimensional simulation. 
    
    The plan of the paper is to discuss the smoothing, and show how it can be formulated so that the linear and angular momentum are conserved in the absence of boundaries and external body forces.  We then derive the Euler equations of motion from a Lagrangian and complete the SPH-$\epsilon$ equations by adding the boundary force and viscous terms.  We discuss the conditions for the kinetic energy to be positive definite and describe the time stepping scheme.  We apply the SPH-$\epsilon$ equations to predict the properties of decaying turbulence initialised by a set of gaussian vortices and  compare them with those  of Clercx et. al. (1999, 2000), and Massen et al. (2002).     
      
 \section{Smoothing}
 \setcounter{equation}{0}
 
  The typical LANS model uses a smoothed velocity $\widehat {\bf v}$ defined in terms of the the disordered velocity $\bf v$ by
  \begin{equation}
  {\widehat{\bf v}}({\bf r}) =  \int  {\bf v}({\bf r}') G(|{\bf r'} - {\bf r}|,\ell) {\bf dr'},
\label{ }
\end{equation}
 where  the integration is over the region occupied by the fluid, and $G $ is a kernel which satisfies
  \begin{equation}
    \int G(|{\bf r'} - {\bf r}|,\ell)  {\bf dr'} = 1,
\label{ }
\end{equation}
and is a member of a sequence of functions which tends to the $\delta$ function in the limit where $\ell \rightarrow 0$. A typical example is a Gaussian, though in practice we use smooth functions that have compact support. The length scale $\ell$ determines the characteristic width of the kernel. For the reader familiar with SPH it is useful to note that the kernel G has the same properties as the kernel W used in SPH,  and $\ell$ has the same significance as the length scale $h$ used with the SPH kernel W.  In the applications to be described here we replace $G$ by the $W$ used in the SPH fluid dynamics and we replace $\ell$ by $h$.

   It is common practice (e.g. Chen et al. (1999), Lunasin et al. 2007) to replace the integral smoothing  by the implicit differential smoothing
\begin{equation}
 {\bf v}({\bf r}) =  ( 1- \alpha^2 \nabla^2)   {\widehat {\bf v}}({\bf r}),
\label{ }
\end{equation}
where $\nabla^2$ is the Laplacian operator, and $\alpha$ has the dimensions of length and determines the length scale below which velocity variations are smoothed.  If $\widehat {\bf v}$ and ${\bf v}$ are expanded in a Fourier series with coefficients of the Fourier term $e^{\imath {\bf k} \cdot {\bf  r}}$, ${\widehat C}_k$ and $C_k$ respectively, then equation (2.1) gives
\begin{equation}
{\widehat C}_k = { \widetilde G}(k)C_k 
\end{equation}
  where  ${ \widetilde G}(k)$ is the Fourier transform of $G$.  If $G$ is a smooth enough function the Fourier transform decreases rapidly with increasing $k$ and ${\widehat C}_k \ll C_k $  The differential smoothing (2.3) behaves similarly and leads to 
 \begin{equation}
{\widehat C}_k = \frac {C_k }{1 + \alpha^2 k^2}.
\end{equation} 
 A disadvantage with (2.3), and with the smoothing discussed by Monaghan (2002) for a preliminary form of SPH turbulence, is that it is implicit and can only be solved by iteration when the domain is complicated and Fourier spectral methods cannot be used. In three dimensions the large number of iterations required may then make the implicit smoothing impracticable. The implicit SPH smoothing (Monaghan 2002) was too slow even  for two dimensional problems because of the large number of iterations required,  and the implicit smoothing (2.3) greatly increases the computational time of LANS-$\alpha$ ocean models (Hecht et al. 2008). We therefore consider a different smoothing which can be converted to the SPH formulation easily, and involves negligible extra work. The smoothing is defined by 
  \begin{equation}
  {\widehat{\bf v}}({\bf r}) =   {\bf v}({\bf r}) + \epsilon \int ({\bf v}({\bf r}') -{\bf v}({\bf r})) G(|{\bf r'} - {\bf r}|,\ell)  {\bf dr'},
\label{ }
\end{equation}
where  $\epsilon$ is a constant and $0<\epsilon <1$. In this case the Fourier coefficients satisfy 
 \begin{equation}
{\widehat C}_k = C_k (1 + \epsilon ( { \widetilde G}(k)-1)),
\end{equation}
and  ${\widehat C}_k \rightarrow (1- \epsilon)C_k$ as $k \rightarrow \infty$.  Provided $(1-\epsilon) \ll1  $ and $0<\epsilon <1$ then ${\widehat C}_k \ll C_k$ as $k \rightarrow \infty$.  The smoothed velocity spectrum therefore retains the same {\it form} as the unsmoothed spectrum, though it is reduced in magnitude.  Of course, in a dynamical calculation, the spectrum of a quantity such as the kinetic energy depends on the smoothing in a more complicated way.  The standard LANS-$\alpha$ model changes the spectrum for high $k$ as in (2.5). In the calculations we describe in this paper, the typical value of $\epsilon$ is 0.8,  so that the high order  Fourier coefficients are reduced by a factor 0.2 every step.  We refer to this model as the SPH-$\epsilon$ turbulence model.

  The SPH equivalent of (2.6)  is the XSPH  equation (Monaghan 1989) which can be obtained by using the SPH procedure for converting integrals into summations over particles. This procedure begins with the integral interpolant which estimates scalar, vector  or tensor function $T$ at any point by $T_I$ defined by
\begin{equation}
  T_I ({\bf r}) =  \int T({\bf r'})W(|{\bf r'} - {\bf r}|,h)  {\bf dr'}, 
 \end{equation}
 where $W$ is a smoothing kernel normalized as for $G$ in (2.2),  and $h$ is a length scale associated with the kernel.  We use kernels with compact support which vanish for points separated by more than $2h$.
 
This integral can be approximated by summing over particles (the errors involved with this are discussed by Monaghan 2005) to give 
\begin{equation}
T({\bf r} )=  \sum_b \frac{m_b }{ \rho_b } T_b W((| {\bf r} - {\bf r}_b |,h),
\end{equation}
where $m_b$, $T_b$ and $\rho_b$ are the mass, value of $T$ and the density of particle $b$.  The subscript $I$ in (2.8) has been dropped for simplicity. An example of this interpolation is the density which can be estimated anywhere by 
\begin{equation}
\rho({\bf r} )=  \sum_b m_b W((| {\bf r} - {\bf r}_b |,h).
\end{equation}
For other examples see Monaghan(2005). The summation is over all particles but the only contributions to the value of $T$ at particle $a$ are those within $2h$. Efficient methods of finding these particles means that the work done to find $T$ for all particles is proportional to the number of particles.  Spatial derivatives of quantities such as $T({\bf r})$ can be obtained at the position of any particle by differentiating (2.9) analytically and evaluating the resulting expression at the position of that particle.

Applying this procedure to (2.6) the smoothed velocity for particle $a$ becomes
\begin{equation}
{\widehat v}_a = {\bf v}_a + \epsilon \sum_b \frac{m_b}{\rho_b} ({\bf v}_b - {\bf v}_a) G_{ab},
\label{ }
\end{equation}
Where  $G_{ab}$ denotes $G(|{\bf r}_a - {\bf r}_b |,\ell)$. However, because the centre of mass should move with constant velocity when the fluid is isolated and free of external forces, we require that
\begin{equation}
\frac{d}{dt} \sum_am_a {\bf r}_a = \sum_a m_a \frac{d {\bf r}_a}{dt} = \sum_a m_a {\widehat v}_a = \sum_a m_a{\bf v}_a = \mathcal {P},
\label{ }
\end{equation}
where $\mathcal{P}$ is the constant momentum.The second equality follows from the fact that elements of fluid move with the smoothed velocity.  The last equality follows from the fact that we use a Lagrangian. However, to satisfy the third equality the summation term in (2.11) must be  anti-symmetric in $a$ and $b$ when multiplied by $m_a$.  To this end we replace $\rho_b$ by a symmetric density $\widetilde{\rho}_{ab}$ and $\ell$ by a symmetric form $\ell_{ab}$, for example  $\ell_{ab} = \frac12(\ell_a + \ell_b)$, and write  
\begin{equation}
\frac{G_{ab} }{\widetilde  \rho_{ab}} = \frac{K(|{\bf r}_a - {\bf r}_b |, \ell_{ab})}{ \ell_{ab}^d \widetilde \rho_{ab}}
\label{ }
\end{equation}
 where $d$ is the number of dimensions.  In compressible gas dynamics $\ell^d_a \propto 1/\rho_a$ which means that $\ell$ is proportional to the local average particle spacing. The same principle applies also to the nearly incompressible fluid where the changes in density and hence $\ell$ are small. We can now define the symmetric density by the condition
 \begin{equation}
 \ell_{ab}^d \widetilde\rho_{ab} = M,
\end{equation}
 where $M$ is a mass which is held constant (suggested by Daniel Price).  This mass is determined by the initial state where $\rho$ and $\ell$ are the same for each particle. The final form of the smoothing is then
 \begin{equation}
{\widehat v}_a = {\bf v}_a + \epsilon \sum_b \frac{m_b}{M} ({\bf v}_b - {\bf v}_a) K_{ab},
\label{ }
\end{equation}

  In the absence of external forces and fixed boundaries the angular momentum is also conserved because $\sum_a m_a {\widehat{\bf v}}_a \times {\bf v}_a = 0$, and then
 \begin{equation}
\frac{d}{dt} \sum_am_a {\bf r}_a \times {\bf v}_a = \sum_a m_a \left ({\widehat{\bf v}}_a \times {\bf v}_a + {\bf r}_a \times \frac{d {\bf v}_a}{dt} \right)  = \sum_a m_a {\bf r}_a \times \frac{d {\bf v}_a}{dt} =0,
\label{ }
\end{equation}
 where the last equality follows from the invariance properties of the SPH-$\epsilon$ equations (see below).

 In addition to ensuring conservation of linear and angular momentum  this choice of $\widetilde{\rho}_{ab}$ means that we do not have to consider the change of the density term in (2.15) when working out the Lagrange equations. A further simplification, suitable for the nearly incompressible fluid, is to assume $\ell_{ab}$ is also constant when working out the Lagrange's equations. 
 %----------------- Lagrange's equations  -----------------------------------------------
 %-------------------------------------------------------------------------------------------------
 \section{ Lagrange's Equations} 
 \setcounter{equation}{0}
 
  The Euler equations for the fluid are the Lagrange equations  obtained from the SPH equivalent of Eckart's Lagrangian (Eckart 1960).
 \begin{equation}
 L = \sum_bm_b \left (  \frac12 {\bf v}_b \cdot {\bf {\widehat v}}_b - u(\rho_b,s_b) \right ),
\label{ }
\end{equation}
where $u(\rho_b,s)$ is the internal energy per unit mass associated with the equation of state. This energy depends, in general, on the density $\rho$ and the entropy per unit mass $s$ which, for the non dissipative Lagrangian formulation,  does not vary with time. In this paper we assume each particle has the same entropy. The invariant energy (in the absence of viscous dissipation and external or boundary forces) is
\begin{equation}
\label{ }
E = \sum_b m_b\left (  \frac12 {\bf v}_b \cdot {\bf {\widehat v}}_b + u(\rho_b,s_b) \right ).
\end{equation}
Lagranges equations for particle $c$ are 
\begin{equation}
\frac{d}{dt} \left ( \frac{\partial L}{\partial  {\widehat {\bf v}_c}} \right) = \frac{\partial L}{\partial {\bf r}_c}.
\end{equation}
The Lagrangian must therefore be written in terms of the $ {\bf {\widehat v}}$  and ${\bf r}$ before the Lagrange equations are worked out.  In order to do this  it is convenient to write (2.15) in the form
\begin{equation}
{\widehat {\bf p}_a} =m_a{\widehat {\bf v}_{a} } = \sum_b D_{ab}{\bf v}_b,
\end{equation}
where 
\begin{equation}
D_{ab} = \delta_{ab} (m_a - \epsilon \sum_{j\ne a} \sigma_{aj} K_{aj} )+ \epsilon(1-\delta_{ab}) \sigma_{ab} K_{ab},
\end{equation}
 $\sigma_{ab} = m_am_b/M$ and $\delta_{ab}$ is the Kronecker delta.  Note that $D_{ab}$ is symmetric and the matrix ${\bf D}$ with elements $D_{ab}$ is a square, symmetric matrix. We can then write 
\begin{equation}
 {\bf v}_{a}  = \sum_b (D)^{-1}_{ab} {\widehat {\bf p}_b},
\end{equation}
where $(D)^{-1}_{ab}$ is the $ab$ component of the matrix inverse to $D$. The kinetic energy 
\begin{equation}
E_K =  \frac12 \sum_a m_a {\widehat {\bf v}}_a \cdot {\bf v}_a,
\end{equation}
can then be written in the form
\begin{equation}
E_K =  \frac12 \sum_a {\widehat {\bf p}}_a \cdot \sum_b (D)^{-1}_{ab} {\widehat {\bf p}_b}.
\end{equation}
Therefore 
\begin{equation}
\frac{\partial E_K}{ \partial \widehat {\bf v}_c } = \frac12 \sum_a\sum_b ( m_a \delta_{ac} {\widehat {\bf p}}_b + m_b \delta_{bc} \widehat {\bf p}_a ) D^{-1}_{ab}.  
\end{equation}
The first and second terms in the summation are identical, and equal to $\frac12 m_c {\bf v}_c$, so that the canonical momentum is $m_c {\bf v}_c$.

The right hand side of Lagrange's equations (3.3) has two contributions, one from  kinetic energy term and one from the elastic energy.  The second leads to the usual SPH pressure equations (see for example Monaghan 1992, 2005).  The first can be written 
\begin{equation}
\frac{\partial E_K}{\partial {\bf r}_c} =  \frac{\partial }{\partial {\bf r}_c} \left ( {\widehat {\bf p} }{\bf D}^{-1} {\widehat {\bf p}} \right )  =   {\widehat {\bf p} } \left (\frac{\partial } {\partial {\bf r}_c}{\bf D}^{-1} \right) {\widehat {\bf p}}.
\end{equation}
where $\widehat {\bf p} = (\widehat {\bf p}_1, \widehat {\bf p}_2, \widehat {\bf p}_3, \cdots) $, and the subscript denotes a particle label.  The row or column form is taken as required for the matrix products.  The derivative $\partial /\partial {\bf r}_c$ is to be interpreted in terms of a separate matrix equation for each component of ${\bf r}_c$. Because ${\bf D}$ is a square symmetric matrix so is its inverse, and we can use the identity
\begin{equation}
\frac{\partial } {\partial {\bf r}_c}{\bf D}^{-1}  = - {\bf D}^{-1} \left ( \frac{\partial } {\partial {\bf r}_c}{\bf D} \right ) {\bf D}^{-1}.
\end{equation}
If this result is substituted into (3.10) and note is taken of (3.4),  we can write (3.10) as 
\begin{equation}
\frac{\partial E_K}{\partial {\bf r}_c}  =    -{\bf v} \left (\frac{\partial {\bf D} } {\partial {\bf r}_c } \right)  {\bf v},
\end{equation}
where ${\bf v} = ({\bf v}_1, {\bf v}_2, {\bf v}_3, \cdots )$ and in (3.12) the column vector form is used.  We can then write 
\begin{equation}
\frac{\partial E_K}{\partial {\bf r}_c}  =    -\sum_a\sum_b{\bf v}_a \cdot {\bf v}_b  \left (\frac{\partial D_{ab} } {\partial {\bf r}_c } \right).
\end{equation}
The expression (3.5)  for $D_{ab}$ contains two summation terms and in each case they involve derivatives of the kernel. However, the discussion after (2.12) shows that we need only consider the explicit dependence on the coordinates, and neglect the variation of $\ell_{ab}$ and $\widetilde \rho_{ab}$.  If the fluid is compressible then the variation of $\ell_{ab}$ involves extra spatial derivatives but in this paper we assume the variations in $\ell_{ab}$ are negligible.  The first contribution to $D_{ab}$ is
\begin{equation}
D^{ (1) }_{ab} =  \delta_{ab} (m_a - \epsilon \sum_{j\ne a} \sigma_{aj} K_{aj}),
\end{equation}
with derivative
\begin{equation}
\frac{\partial D_{ab}^{(1)} } {\partial {\bf r}_c }  = - \epsilon \delta_{ab} \sum_{j \ne a} \sigma_{aj} \frac{\partial K_{aj}}{\partial {\bf r}_a} (\delta_{ac} - \delta_{jc}).
\end{equation}
The contribution to the derivative of the kinetic energy is 
\begin{equation}
\frac{ \partial E_K^{(1)} } {\partial {\bf r}_c }  = \frac12 \epsilon \sum_a\sum_b {\bf  v}_a \cdot  {\bf v}_b \delta_{ab} \sum_{j\ne a} \sigma_{aj }\frac{\partial K_{aj}}{\partial {\bf r}_a} (\delta_{ac} - \delta_{jc}).
\end{equation}
If this expression is simplified, noting that $\partial K_{ab}/\partial {\bf r}_a = - \partial K_{ab}/\partial {\bf r}_b$, it becomes
\begin{equation}
\frac{ \partial E_K^{(1)} } {\partial {\bf r}_c }  = \frac12 \epsilon \sum_{a \ne c} (v_a^2 + v_c^2)\sigma_{ac }\frac{\partial K_{ac} }{\partial {\bf r}_c}.
\end{equation}

The second contribution to $D_{ab}$ is
\begin{equation}
D^{ (2) }_{ab} =  \epsilon (1- \delta_{ab}) \sigma_{ab} K_{ab}.
\end{equation}
with derivative
\begin{equation}
\frac{\partial D_{ab}^{(2)} } {\partial {\bf r}_c }  =  \epsilon (1- \delta_{ab})  \sigma_{ab} \frac{\partial K_{ab}}{\partial {\bf r}_a} (\delta_{ac} - \delta_{bc}),
\end{equation}
and the contribution to the derivative of the kinetic energy is 
\begin{equation}
\frac{ \partial E_K^{(2)} } {\partial {\bf r}_c }  =  -\frac12 \epsilon \sum_a\sum_b  {\bf v}_a \cdot  {\bf v_b} (1- \delta_{ab} ) \sigma_{ab} \frac{\partial K_{ab}}{\partial {\bf r}_a} (\delta_{ac} - \delta_{bc}).
\end{equation}
If this expression is simplified it can be written
\begin{equation}
\frac{ \partial E_K^{(2)} } {\partial {\bf r}_c }  = -\epsilon \sum_{a \ne c} \sigma_{ac} {\bf v}_a\cdot {\bf v}_c \frac{\partial K_{ac}}{\partial {\bf r}_c}.
\end{equation}
Combining (3.17) and (3.21) gives
\begin{equation}
\frac{ \partial E_K } {\partial {\bf r}_c }  =  \frac12 \epsilon \sum_{a \ne c} v_{ab}^2\sigma_{ac }\frac{\partial K_{ac} }{\partial {\bf r}_c},
\end{equation}
where ${\bf v}_{ab} = {\bf v}_a - {\bf v}_b$.

The contribution to the equations of motion from the internal energy term can be calculated easily. The details can be found in Monaghan (2005), but the essentials are as follows.  If there is no dissipation the first law of thermodynamics for unit mass is  $du = dp/ \rho^2$.  For the slightly compressible fluid we assume 
\begin{equation}
P = \frac{\rho_0 c_s^2 }{\gamma} \left ( \left (\frac{\rho}{\rho_0} \right )^\gamma -  1 \right ) ,
\end{equation}
where $\rho_0$ is a reference density, $c_s$ is the speed of sound, and $\gamma = 7$ in this paper.  For the weakly compressible case we choose $c_s = 10 V_{max}$ where $V_{max}$ is the maximum fluid velocity, so that the density fluctuation $\delta \rho/\rho \sim (v_{max}/c_s)^2 = 0.01$. If  the density is estimated by the SPH summation  
\begin{equation}
\rho_c = \sum_b m_b W_{cb},
\end{equation}
the contribution to the right hand side of Lagrange's equation for particle $c$
\begin{equation}
-\sum_bm_b \frac{du_b}{d\rho_b} \frac{\partial \rho_b}{\partial {\bf r}_c}.
\end{equation}
becomes
\begin{equation}
- \sum_b m_b \left (  \frac{P_c}{\rho_c^2} +  \frac{P_b}{\rho_b^2}  \right ) \nabla_c W_{cb}.
\end{equation}
Combining the previous results,  Lagrange's equation for particle $c$ are
\begin{equation}
\frac{d {\bf v}_c}{ dt} = - \sum_b m_b \left (  \frac{P_c}{\rho_c^2} +  \frac{P_b}{\rho_b^2}  \right ) \nabla_cW_{cb}
+ \frac{\epsilon}{2}\sum_b \frac{m_b}{M} v_{cb}^2 \nabla_cK_{cb}.
\end{equation}

This equation is the SPH-$\epsilon$ Euler equation. It is Galilean invariant and invariant to rotations of the coordinate system. For these reasons the linear and angular momentum are conserved.  Because the Lagrangian is not an explicit function of time the energy (3.2) is also conserved. The equations also satisfy a discrete version of Kelvin's circulation theorem (this follows from the same argument used by Monaghan (2005)). The term involving  $\epsilon$ in (3.27)  is related to the velocity derivative terms in the LANS-alpha equations. 

In this paper we choose the smoothing function $G$ to be the same as $W$ and $\ell$ to be $h$.  In that case we can replace (3.27) by
\begin{equation}
\frac{d {\bf v}_c}{ dt} = - \sum_b m_b \left (  \frac{P_c}{\rho_c^2} +  \frac{P_b}{\rho_b^2}  -\frac{\epsilon}{2} \frac{v_{bc}^2}{\widetilde \rho_{bc} } \right ) \nabla_cW_{cb},
\end{equation}
 where $h$ is allowed to vary according to $h_a \propto 1/\rho_a^{1/2} $ This   variation is $<  1\%$ when the speed of sound is calculated as described after (3.23).  

 In addition to the acceleration equation just derived we use a density continuity equation for particle $c$ which may be deduced directly from the continuum continuity equation, or  by differentiating (3.24) with respect to time. It is
\begin{equation}
\frac{d \rho_c}{dt} =  \sum_b m_b {\widehat {\bf v}}_{bc} \cdot  \nabla_c W_{bc},
\end{equation}
where ${\widehat {\bf v}}_{bc} = {\widehat {\bf v}}_b - {\widehat {\bf v}}_c$, and 
\begin{equation}
\frac{d {\bf r}_c}{dt} =  {\widehat {\bf v}}_c.
\end{equation}
The acceleration equation (3.27) can also be derived using the SPH continuity equation as a constraint equation when deducing Lagrange's equations from the Least Action variational principle.

% ----------viscous equations  and boundary forces----------------

\subsection{ Boundaries and boundary forces}

In this paper the boundaries are defined by boundary force particles which exert forces on the fluid. This technique is similar to the immersed boundary method of Peskin (1957, 2002) and is discussed in detail by Monaghan and Kajtar (2009).  The basic idea is that if the interaction between the fluid and boundary particles is sufficiently smooth, and the forces between the particles is along their line of centres, the force on a fluid particle is normal to the boundary high accuracy.  Monaghan and Kajtar (2009) showed that this is true for both curved and planar surfaces, provided the boundary force particles were spaced less than half the spacing of the fluid particles. 

 When the boundary force particles are included (3.28) becomes
 \begin{equation}
\frac{d {\bf v}_c}{ dt} = - \sum_b m_b \left (  \frac{P_c}{\rho_c^2} +  \frac{P_b}{\rho_b^2}  -\frac{\epsilon}{2} \frac{v_{bc}^2}{\widetilde \rho_{bc} } \right ) \nabla_cW_{cb} + \sum_{k=1}^{N_b} \sum_{j\in S_k} {\bf f}_{c j}  ,
\end{equation}
where the sum over $k$ is over the $N_b$ bodies, and $j\in S_k$ denotes the boundary particles of body $k$. The details of the boundary force are given in the appendix.  Because the boundary force between a pair of particles is radial it is possible to write it in terms of a pair potential which can then be included in the Lagrangian.  This boundary force does not change the conservation properties of the Lagrangian.

The present formulation is general and allows for an arbitrary number of boundaries which could include immersed bodies.  These bodies may move either by specifying their motion or by determining it from the forces exerted on the boundary force particles by the fluid particles.  In this paper the only boundary is the fixed square boundary containing the fluid.

\subsection{Viscous forces}

To complete our model we need to add  a viscous term.  There are several SPH approximations to the Navier-Stokes viscosity (see Monaghan 2005 for two examples).  The viscous equations take the form
 \begin{equation}
\frac{d {\bf v}_c}{ dt} = - \sum_b m_b \left (  \frac{P_c}{\rho_c^2} +  \frac{P_b}{\rho_b^2} + \Pi_{bc}  -\frac{\epsilon}{2} \frac{v_{bc}^2}{\widetilde \rho_{bc} } \right ) \nabla_cW_{cb} + \sum_{k=1}^{N_b} \sum_{j\in S_k } m_j\left ( {\bf f}_{c j} - \Pi_{jc} \nabla_c W_{bc} \right ) .
\end{equation}
where
\begin{equation}
\Pi_{bc}  = - \frac{ \alpha v_{sig} {\bf v}_{bc} \cdot {\bf r}_{bc} }{ \bar {\rho}_{bc} r_{bc}}.
\end{equation}
In this expression $\alpha$ is a constant, $v_{sig}$ is a signal velocity which, in this paper, is taken as the speed of sound in the equation of state, and $\bar {\rho}_{bc} = (\rho_b + \rho_c)/2 $. If the density is constant, and the kernel $W$ is a Wendland fourth order kernel (see \S3.3),  it is possible to show that the kinematic viscosity $\nu$ is then given by
\begin{equation}
 \nu = \frac18 \alpha c_s h.
\end{equation}
This result is obtained by converting the SPH summations to integrals (Monaghan 2005). 

%-----------positive definite kinetic energy

\subsection{ Constraints on $\epsilon$ from the kinetic energy}
The kinetic energy should be positive definite.  From (3.4) and (3.7) the kinetic energy is 
\begin{equation}
E_K = \frac12 \sum_a\sum_b {\bf v}_a D_{ab} {\bf v}_b,
\end{equation} 
which can also be written as
\begin{equation}
E_K = \frac12 \sum_a m_a v_a^2 - \frac12 \epsilon \sum_a\sum_b\frac{m_a m_b}{M} v_{ab}^2 K_{ab},
\end{equation} 
which shows that $E_K$ is less than the kinetic energy using the unsmoothed velocity.  This is to be expected because the smoothing reduces the velocity associated with short length scales. However, it is important to ensure that $E_K$ is positive definite otherwise the energy might be reduced by increasing the velocity.  Clearly this condition depends on the magnitude of $\epsilon$.  In the calculations described in this paper where $0 < \epsilon <0.9$ we always find $E_K > 0$. 

  A sufficient condition on $\epsilon$ can be found by determining the sufficient condition for  the eigenvalues of the matrix $\bf D$ to be positive. This can be done using Gerschgorin's theorem. This simple theorem is obtained as follows.  Let ${\bf D} {\bf q} = \lambda {\bf q}$. We normalize the eigenvector $\bf q$ by making the largest component $q_k=1$.  Then, from  row $k$ of the matrix ${\bf D}$, we deduce
\begin{equation}
\sum_{j\ne k}  D_{kj}q_j +D_{kk} = \lambda,
\end{equation}
and
\begin{equation}
| \lambda - D_{kk}| \le \sum_{j\ne k} |D_{kj}| |q_j| \le \sum_{j \ne k} |D_{kj}|.
\end{equation}
From this inequality we infer that $\lambda \ge 0$ provided 
\begin{equation}
D_{kk}  > \sum_{j \ne k} |D_{kj}|,
\end{equation}
or 
\begin{equation}
1 - \epsilon \sum_{J \ne k} \frac{m}{M} K_{kj} > \epsilon \sum_{j\ne k} \frac{m}{M} K_{kj}
\end{equation}
To estimate the values of $\epsilon$ for which this inequality is true we consider equal mass SPH particles placed on the vertices of a grid of squares where the side of each square has length $\Delta$.   In the calculations to be described here the smoothing function  is the fourth order Wendland kernel for two dimensions (the order here refers to the way it approaches zero) which has the form
\begin{equation}
G(z,\ell) = \frac{7}{ 64 \pi \ell^2} (2-z/\ell)^4 (1+2z/\ell),
\end{equation}
if $z\le 2 \ell$ and zero otherwise. The smoothing is therefore over a length $2 \ell$. If $\ell = \Delta$ the sufficient condition is that $\epsilon < 1.04$. If $\ell = 1.5 \Delta$ the conditions is $\epsilon  < 0.667$.  If $\ell $ is sufficiently large relative to $\Delta$ we can replace the summations by an integration over the area after subtracting off the term with $j = k$ in (3.6). We then find that the sufficient condition is 
$1 > 2 \epsilon (1-  7 \Delta^2/(4 \pi \ell^2))$ or, when $\ell/\Delta \rightarrow \infty$, the sufficient condition becomes  $\epsilon < 0.5$.  However, as mentioned earlier, these sufficient conditions appear to be excessively pessimistic because a wide variety of  calculations with $\epsilon = 0.9$ have positive $E_K$.

%--------------------------------------------
% time stepping
%--------------------------------------------
\subsection{Time Stepping}

The SPH turbulence equations were integrated using a time stepping scheme  that is second order and based on the Verlet symplectic method though it is more complicated because the force depends on the velocity.  The ideal form is reversible in the absence of viscosity.  For convenience we write the equation in the form
\begin{eqnarray}
\frac{d {\bf v}_c}{dt}  & = & {\bf F}({\bf r},\rho,{\bf v})_c, \\
\frac{d \rho_c}{dt}  & = & B({\bf r},{\widehat {\bf v} }),\\ 
\frac{d{\bf r}_c}{dt} & = &{\widehat {\bf v}}_c,\\
{\widehat {\bf v} }_c & =  &{\bf g}({\bf v}_c,{\bf r}_c),
\end{eqnarray}
We use the notation that $A^0$ denotes a quantity $A$ at the beginning of the current step, $A^{1/2}$  at the midpoint of the step, and $A^1$ at the end of the step.  The time stepping equations, where $\delta t$ is the time step,  can then be written
\begin{eqnarray}
{\bf r}_c^{1/2} & =  &{\bf r}_c^0 + \frac12 \delta t {\widehat {\bf v}}^0,\\
\rho^{1/2}_c  & = & \rho^0_c + \frac 12 \delta t B^0_c.
\end{eqnarray}
 The time step can then be completed by first calculating ${\bf v}_c^1$  according to
  \begin{equation}
{\bf v}_c^1  =  {\bf v}_c^0 + \delta t {\bf F}_c( {\bf r}^{1/2},\rho^{1/2},  ( {\bf v}^1 + {\bf v^0})/2),
\end{equation}
after which ${\widehat {\bf v}}^1$ can be calculated according to
\begin{eqnarray}
{\widehat {\bf v}^1 } & =  &{\bf g}({\bf v}^1,{\bf r}^1),\\
{\bf r}_c^{1} & =  & {\bf r}_c^{1/2} +\frac12 \delta t {\widehat {\bf v}^1 }_c .
\end{eqnarray}
This algorithm is reversible in the absence of dissipation. However, iteration is required to solve (3.48).  We therefore compromise by replacing $( {\bf v}_c^1 + {\bf v^0}_c)/2$ with  ${\bf v}_c^{1/2}$ and estimate ${\bf v}_c^{1/2}$  by
\begin{equation}
{\bf v}_c^{1/2} = {\bf v}_c^0 + \frac12 \delta t {\bf F}_c^{-1/2}.
\end{equation}
   The  two equations (3.49) and (3.50) also require iteration. Clearly the number of iterations depends on the time step and $\epsilon$ and, if either  is too large, the iterations will not converge. The condition on the time step to guarantee the iteration will converge is easily worked out. However, for all the calculations in this paper where $0<\epsilon <0.9$, the iterations converge in 3 steps with an error of $10^{-6} $ with just the CFL condition $\delta t < 0.5h/c_s$.  Note that the function $\bf g({\bf v},\bf {r})$ also contains $\ell$ or, in the present case, $h$.  When iterating (3.49) and (3.50)  this is replaced by the midpoint value calculated using the midpoint density.  
 %---------------------------------------------------------------------------------
 \section{Applications to turbulence in a square 2D rigid box}
 %----------------------------------------------------------------------------------
 \setcounter{equation}{0}
 We consider a two dimensional fluid inside  a rigid square with side length 1m and  no-slip  boundary conditions. This problem  provides a good test of turbulence models because it has been studied in great detail by laboratory experiment and by highly accurate numerical simulations (see for example Clercx et al. 1999, Clercx and Heijst 2000, and  Maassen et al. 2002).  Furthermore, the computer time is greatly reduced relative to three dimensions and the calculations  described here were run on a MacBook Pro.  The results of Clercx et al. which we use for comparison have Reynolds numbers $\Re$ in  the range 1000 to 5000. In this range the results are similar. 
 
  %----------------------------------------------------------------------
 %----------------------------------------------------------------------
 \begin{figure}[h!]
 \begin{center}
\includegraphics[width=0.7 \textwidth ]{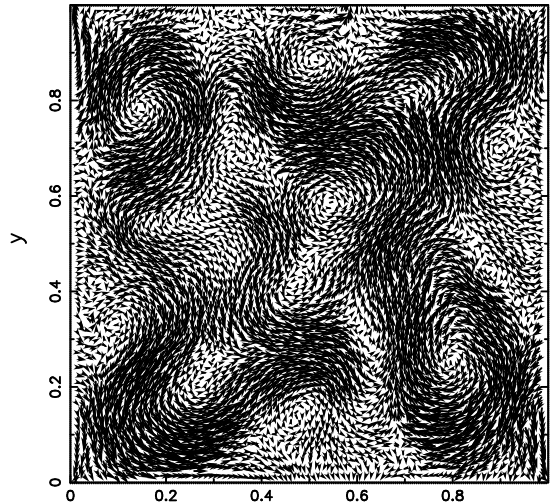}
\caption{ {\small The ${\widehat {\bf v}}$  field  for $\epsilon = 0.9$ shown after 5s ($\sim$1 typical time scale). The small length scale structures have been replaced by larger scales.} }
\end{center}
\end{figure}
 %------------------------------------------------------------------------
 %------------------------------------------------------------------------

  For the SPH simulation we use the 4th order  Wendland kernel $W(r,h)$ identical to $G$ defined by  (3.41). The initial particle density was  $1.01\rho_w $ where $\rho_w$ is the density of water,  and the reference density for the equation of state. This produces a small background pressure.  $h$ was set to  $1.5 dp$ where $dp$ is the particle spacing.  The smoothing of the velocity has $\ell=h$ so that the smoothing is over a circle of radius $\sim 3dp$ around any specified particle.  
  
  The Reynolds number $\Re$, with velocity 0.12 m/s (which is the typical maximum velocity in the box after the few seconds),  and length scale equal to half the side of the box, was 1000. This $\Re$  is similar to those used by Clercx, Maassen and van Heijst (1999) who considered $\Re$ in the range  $500< \Re <2000$.  With the $\Re$ specified, and the kinematic viscosity given by (3.34), the value of $\alpha$ for a specified resolution and therefore specified $h$, can be calculated.  The resolution length for a well resolved calculation can be estimated in the same way as did Clercx, Maassen and van Heijst (1999) from the enstrophy dissipation rate $\zeta$ using the combination $(\zeta/\nu^3)^{1/6} < < ({\rm box  length})/N$ where $N$ is the number of Chebyshev modes along a coordinate axis.  They conclude that for $\Re = 1000$  that $N$ should be $\sim 180$, or the total number of modes is $180^2$. We find that convergence is achieved for an SPH calculation with $150^2$  particles.   
  
   The SPH particles were placed initially on a grid of squares then damped to equilibrium, after which they were set in motion with velocities  specified by a 4 x 4 set of vortices.  A similar  procedure was used by Maassen et al. 2002 though they had higher resolution and used a 10 x 10 set of Gaussian vortices. The vortices were equi-spaced, with spacing 0.2, then given a random shift in both $x$ and $y$ of 0.02 $\eta$ where $\eta$ is a uniformly distributed random number with  $-1< \eta <1$.  The sign of the rotation and hence vorticity due to each vortex was $\pm$ in a chess board pattern.  The velocity at $\bf r$ due to a vortex at ${\bf R}$ is 
 \begin{equation}
{\bf v}({\bf r}) = \Omega {\bf e}_z \times ({\bf r} - {\bf R}),
 \label{ }
\end{equation}
 where ${\bf e}_z $ is a unit vector normal to the fluid and $\Omega$ is given by
 \begin{equation}
 \Omega = \frac{d}{2 \pi |{\bf r} - {\bf R}|^2} \left (  1 -e^{-( |{\bf r} - {\bf R}|^2/d^2)} \right ),
\label{ }
\end{equation}
and $d = 0.02$. 
 
  %----------------------------------------------------------------------
 \begin{figure}[h!]
 \begin{center}
\includegraphics[width=0.7 \textwidth ]{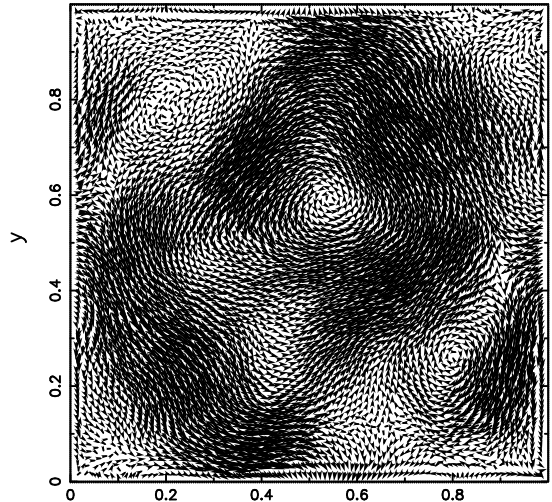}
\caption{ {\small The ${\widehat {\bf v}}$  field  for $\epsilon = 0.9$ shown after 15s (3 typical time scales). The small length scale structures have been replaced by larger scales.} }
\end{center}
\end{figure}
 %------------------------------------------------------------------------

The Fourier transform of the vorticity produced by this velocity field decreases exponentially according to $e^{-(kd/2)^2}$. The no-slip condition requires that the fluid velocity vanishes on the boundary. This was achieved by smoothing the velocity near the boundaries in a similar manner to Clercx et al. (2000). The mean square speed was initially 0.15 giving a typical time scale of $0.5/.15 \sim 3$.
  
  In Fig.1 we show the smoothed velocity field of a  set of fluid particles  with  $\epsilon = 0.9$ and initial spacing 1/75 at a time 5s (or $\sim 1$ time scales) after damping.  Figure 2 shows the smoothed velocity field after 15s (or 5 time scales.  Already much of the complex short length scale velocity field has transformed into larger scales which is similar to that found in experiments and in other simulations (Maassen et al. 2002, their figure 3).  A similar result was found if the simulation is run without smoothing indicating that it is not due to the smoothing but to the normal processes that occur in two dimensional turbulence.
%----------------------------------------------------------------------------------------
 \subsection{ Decay of kinetic energy and enstrophy with time} 
  %----------------------------------------------------------------------
\begin{figure}[h!]
 \begin{center}
\includegraphics[width=0.7 \textwidth ]{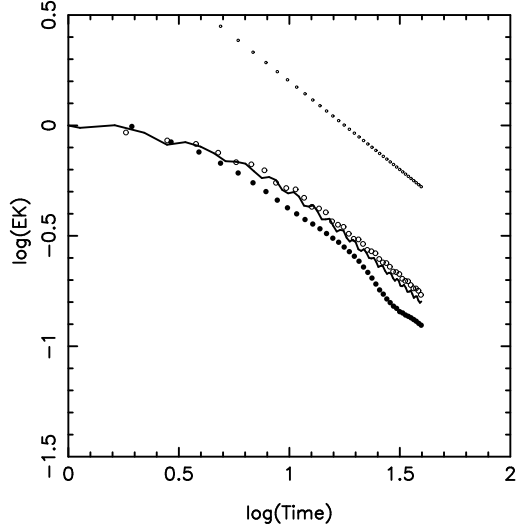}
\caption{ {\small The decay of kinetic energy with time for a simulation with $\epsilon =0$ is shown. The filled circles are for resolution $1/75$, the larger open circle are for resolution $1/125$ and the continuous curve for resolution $1/150$.  The  small open circles  mark a line with time variation $1/t^{0.8}$}.}
\end{center}
\end{figure}
 %------------------------------------------------------------------------
 %----------------------------------------------------------------------
 \begin{figure}
 \begin{center}
\includegraphics[width= 0.7 \textwidth]{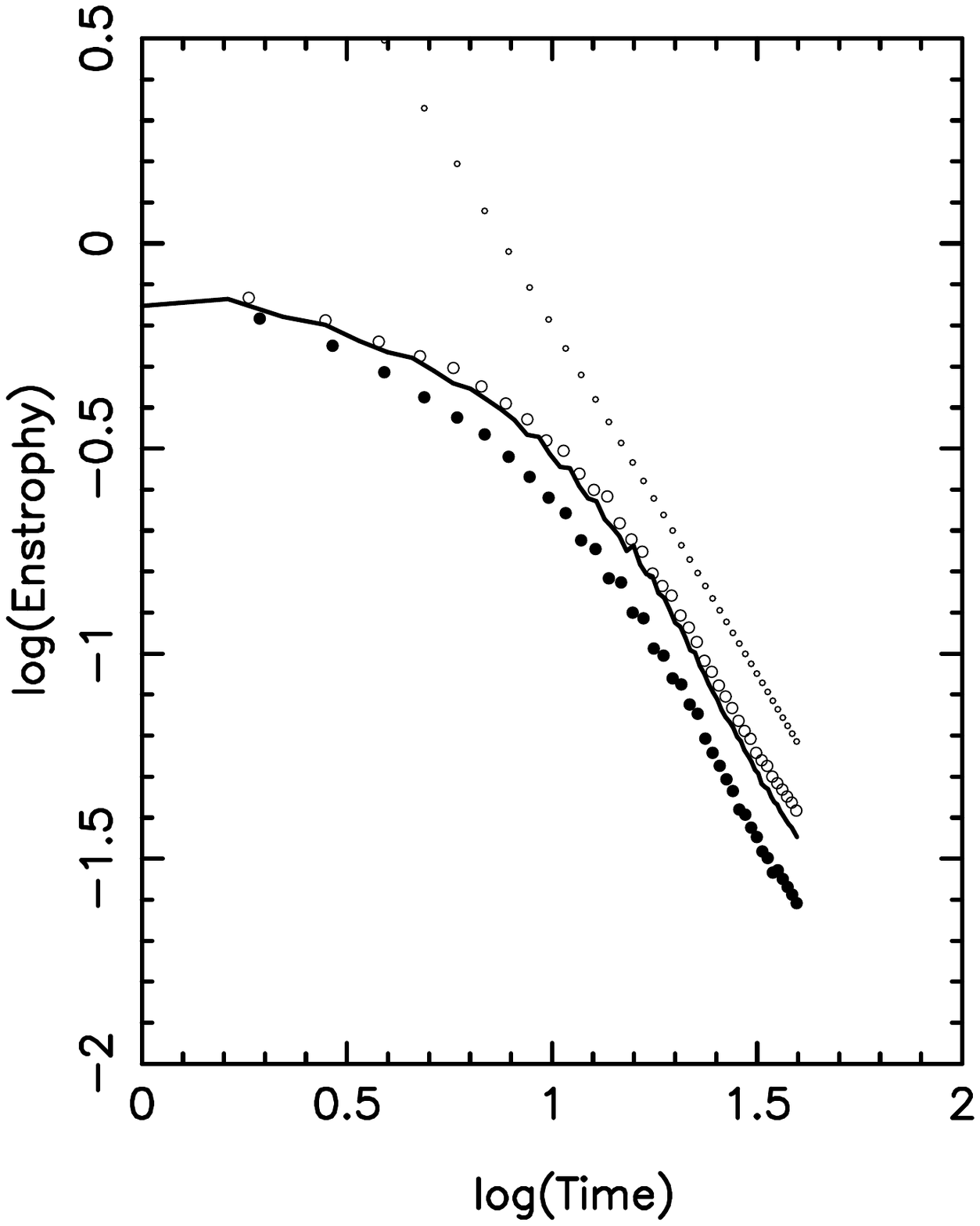}
\caption{{\small The decay of Enstrophy with time with $\epsilon = 0$. The filled circles are for resolution $1/75$, the larger open circle are for resolution 1/125 and the continuous curve for resolution $1/150$.  The small open circles  mark a line with time variation $1/t^{1.7}$}.}
\end{center}
\end{figure}
 %------------------------------------------------------------------------
%----------------------------------------------------------------------
 \begin{figure}
 \begin{center}
\includegraphics[width= 0.7 \textwidth]{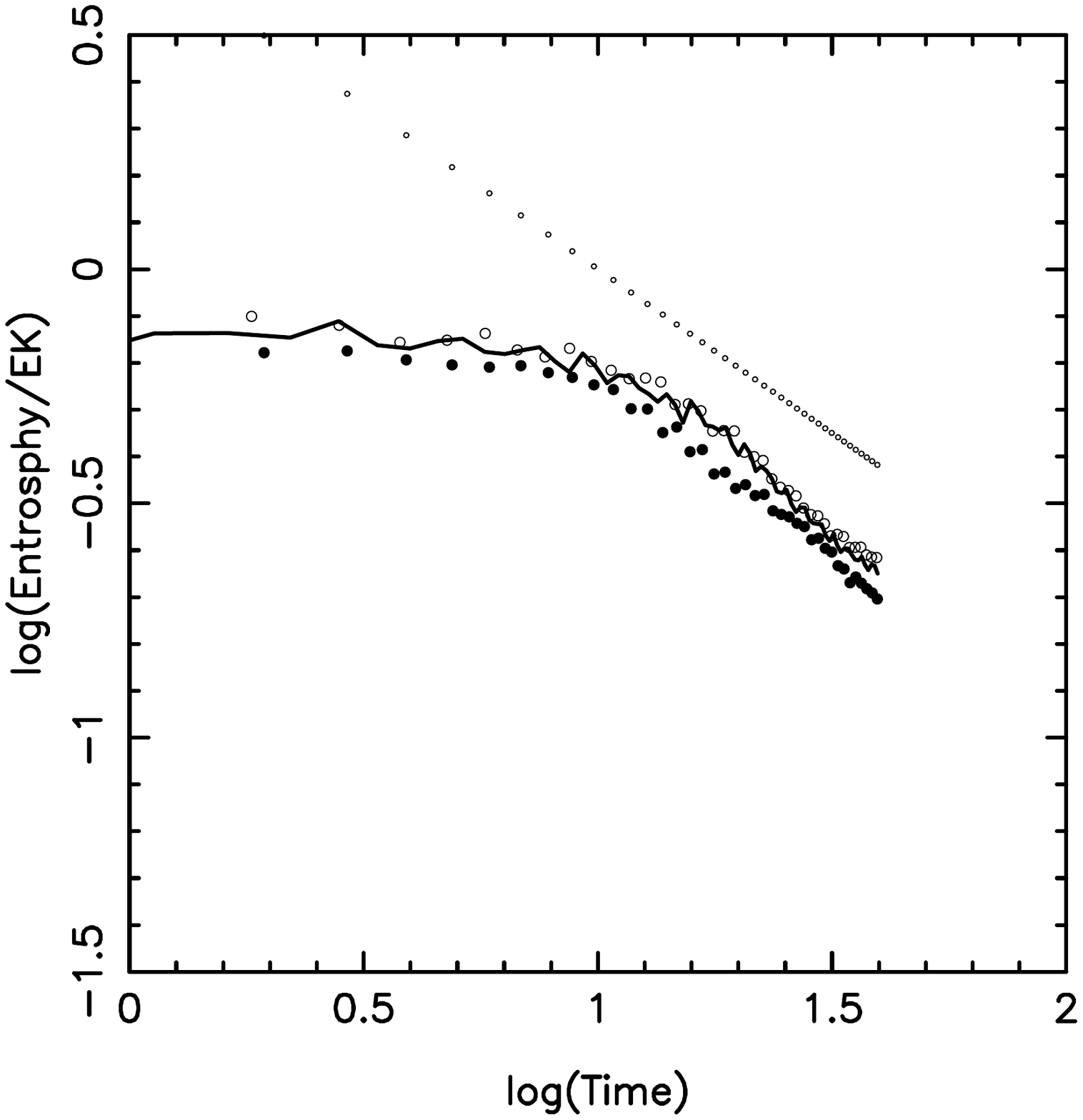}
\caption{{\small The decay of $Enstrophy/E_K$  with time with $\epsilon =0$. The filled circles are for resolution $1/75$, the larger open circle are for resolution 1/125 and the continuous curve for resolution $1/150$.  The small open circles  mark a line with time variation $1/t^{0.7}$}.}
\end{center}
\end{figure}

 In Fig. 3 we show the decay of kinetic energy $E_K$ (defined as in (3.7)), for three simulations with initial particle spacing 1/75, 1/125 and 1/150 and$\epsilon =0$. The decay is similar to the experimental results found by Maassen et al. (2002) for weak stratification (their figure 9(a)). The convergence of the numerical solution is indicated by the closeness of the results for the particle spacing 1/125 and 1/150.  We estimate the error in a simulation with particle spacing $dp$ to be $\propto (dp)^{1.6}$.  Similar results were found for the enstrophy in the same simulation shown in figure 4 where the line of small open circles has the time variation $1/t^{1.7}$ in good agreement with the results of Maassen et al. (2002) (their figure 9(b)). In figure 5 the time variation of the ratio of enstrophy to kinetic energy is shown. This is also in good agreement with  Maassen et al. (2002) (see their figure 17).  Figure 6 shows the decay of the enstrophy with time when $\epsilon = 0.9$.   
 
      %--------------------------------------------------------------------------------------
 \begin{figure}
 \begin{center}
\includegraphics[width= 0.7\textwidth]{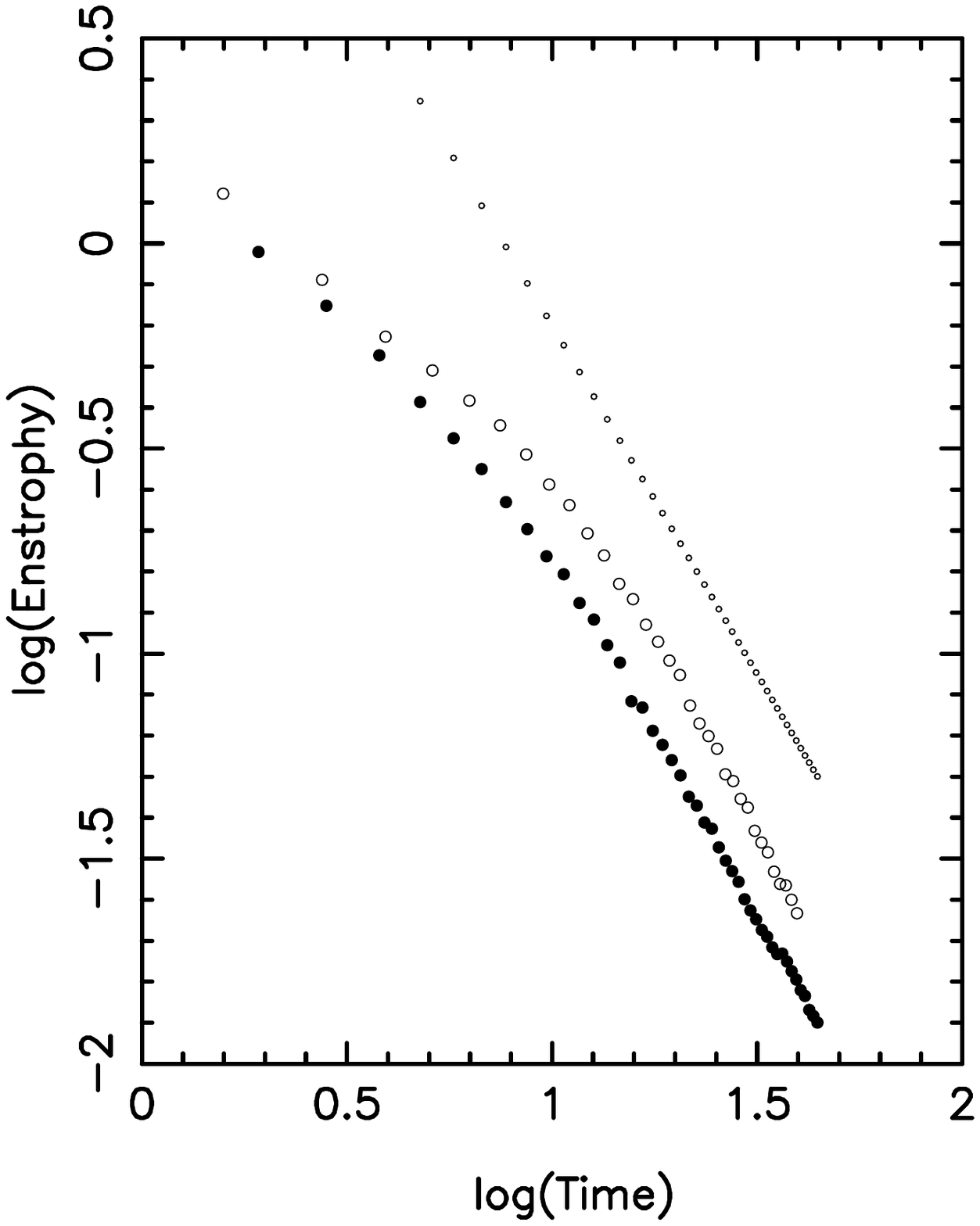}
\caption{ \small The decay of the entrosphy for initial spacing $1/75$ (filled circles), and $1/125$ (large open circles),  with  $\epsilon = 0.9$. The line of small open circles marks $1/t^{1.7}$}
\end{center}
\end{figure}
   %------------------------------------------------------------------------------- 
  \subsection{The second order velocity correlation function}
  %--------------------------------------------------------------------------------
    
 The second order, longitudinal,  velocity correlation function $C_2(R)$  is defined by
 \begin{equation}
C_2(R) = \frac{ \int \int \left [({\bf v}({\bf r}) - {\bf v}({\bf r}')) \cdot {\bf q}/q \right ]^2\delta (q-R) {\bf dr} {\bf dr}'}{  \int \int \delta (r-R) {\bf dr} {\bf dr}' },
\label{ }
\end{equation}
where ${\bf q}  = {\bf r} - {\bf r}'$,  $\delta(x)$ denotes a one dimensional Delta function, and ${\bf dr}$ denotes a volume (area in 2D) element.  Depending on the case we consider the velocity may be the smoothed or unsmoothed velocity. We evaluate the previous expression using SPH summations.  The most efficient way of evaluating these summations is by binning.  We consider every pair of particles. For particles  $a$ and $b$ we calculate an integer $k= Int(r_{ab}/\Delta)$ where $Int$ is the Fortran integer function and $\Delta$ is a suitable small fraction of the maximum value of $R$. The value of 
$ (({\bf v}_a - {\bf v}_b) \cdot {\bf r}_{ab}/r_{ab} )^2$ is added to $ F_k$ and 1 is added to $N_k$. The value of $C_2(R_k)$ is then the final value of $F_k/N_k$ where $R_k = k \Delta$.  Because the system is not homogeneous or isotropic the correlation function was calculated for points  within the square $0.3<x< 0.7$, and $0.3 < y < 0.7$ so that the influence of the boundary was small. 

%------------------------------------------------------------------------------ 
  \begin{figure}
  \begin{center}
\includegraphics[width= 0.7 \textwidth]{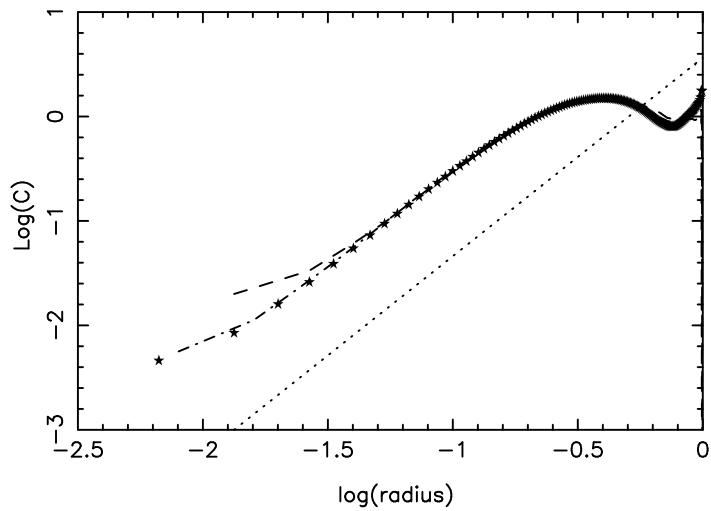}
\caption{{\small The second order velocity correlation functions for three simulations with $\epsilon = 0$, approximately 20s (four  typical time scales) after the velocity was initialized. The dashed line shows the results with initial spacing $1/75$, the dot-dashed results with initial spacing 1/125 and the filled star symbols the results with initial spacing 1/150. The straight  dotted line marks the line $R^{1.9}$}. }
\end{center}
\end{figure}
%--------------------------------------------------------------------------

  An example of the results found using this method are shown in Fig. 7 for the case of no smoothing and initial particle spacings of 1/75, 1/125 and 1/150 at  a time of 20s after the velocities were initialised.  The values of $C_2$ were scaled by the average of $C_2$.  All three correlation functions agree closely until a radius of approximately $3/75$ which is close to $2h$ for the lowest resolution. At this length scale the correlation function with resolution 1/75 is larger, indicating more disorder than for the higher resolutions.   The good agreement between the results for initial particle separation 1/125 and 1/150 over the whole domain confirms the results for the decay of kinetic energy and enstrophy. The radial variation of the correlation function is given approximately by $R^{1.9}$ with a distortion at large $R$ caused by the finite domain sampled, and possibly to the no-slip boundary conditions. The Batchelor theory (Batchelor 1969) of the direct cascade of Enstrophy in two dimensional turbulence  predicts that this second order correlation function should vary with $R$ as $R^2$.   
  %----------------------------------------------------------------------
 \begin{figure}[h!]
 \begin{center}
\includegraphics[width=0.8 \textwidth ]{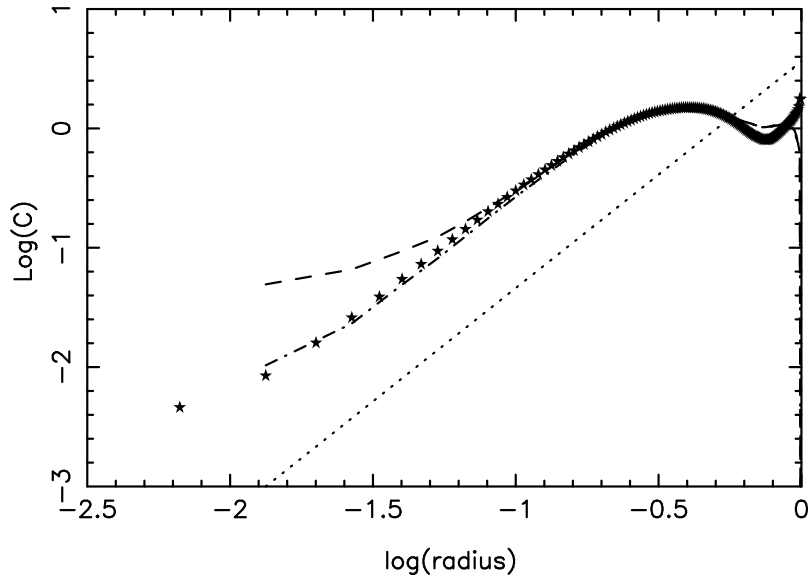}
\caption{{\small The second order velocity correlation function for simulations with initial particle spacing 1/75 and 1/150 at a times 20s after the velocities have been initialized.  The dashed line shows the second order correlation function for the  ${\bf v}$, and the dash-dot line shows the correlation function for ${\widehat {\bf v}}$ in a simulation where $\epsilon = 0.75$.  Note that the correlation function with the smoothed velocity,  and particle spacing 1/75 , is very close to that with no smoothing and spacing 1/150.}}
\end{center}
\end{figure}

     In Figure 8 we show the second order velocity correlation function calculated for both the smoothed and unsmoothed velocities when the initial particle spacing is 1/75 and $\epsilon = 0.75$, and a higher resolution calculation with spacing 1/150 and no smoothing. The star symbols show the results when the initial particle spacing is 1/150 . These results show that with $\epsilon = 0.75$ the results for the lower resolution are in very close agreement with those for the higher resolution, indicating that the smoothed coarse simulation gives a similar velocity field to the higher resolution calculation.  It is interesting  to note that the coarse resolution with no smoothing gives a velocity correlation function between that for $v$ and $\widehat v$ calculated using smoothing.
 
 %----------------------------------------------------------------------
 \begin{figure}[h!]
 \begin{center}
\includegraphics[width=0.7 \textwidth ]{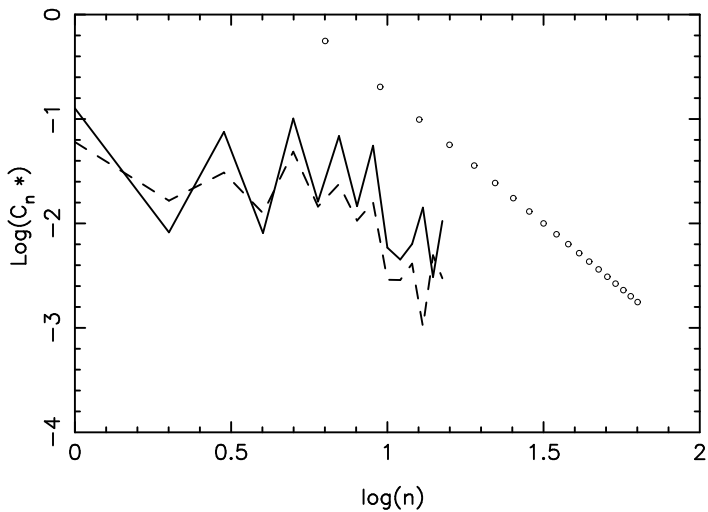}
\caption{{\small The Chebychev spectrum of the kinetic energy at a time 40s after damping with initial particle spacing  1/75. The continuous  line is for $\epsilon = 0$ and the dashed line is for $\epsilon = 0.9$.  Note that the larger $\epsilon$ results in a significant drop in the spectrum after mode number $\sim 8$.  The small circles mark the line $1/n^{2.5}$}.  The zig zag shape shows 
that the odd modes have more energy than the even modes.}
\end{center}
\end{figure}
 %------------------------------------------------------------------------

  %-------------------------------------------------------------------------------------
  
  \subsection{Chebyshev spectrum}
  
      Because the motion in a rigid no-slip box is neither homogeneous nor isotropic Clercx and Heijst (2000) calculated one dimensional spectra both near the boundaries and in the central regions. The spectrum shown in figure 9 was calculated following the procedure used by Clercx et al 1999, where the kinetic energy is expanded in Chebychev polynomials. The version of their procedure adopted here is the following.  We use Chebyshev polynomials $T_n^\star(x) $  related to the standard Chebyshev polynomials $T_n$ by $T^\star_n(x) = T_n(2x-1)$ where $0 \le x \le 1$.  If $E_K$ (defined by (3.7) ) is expanded in these polynomials it can be written
   \begin{equation}
E_K(x,y) = \sum_i\sum_j C_{ij} T^\star_i(x) T^\star_j (y).
\end{equation}
The coefficients $C_{ij}$ were calculated by integrations which were evaluated using SPH summations.  Comparison with the expansion of test functions showed that to estimate the coefficient $C^*_{ij}$ to within $ 10 \%$ required both $i$ and $j$ to be $ < {\rm (box width)}/(5dp)$  where $dp$ is the initial particle spacing.
%----------------------------------------------------------------------
 \begin{figure}[h!]
 \begin{center}
\includegraphics[width=0.7 \textwidth ]{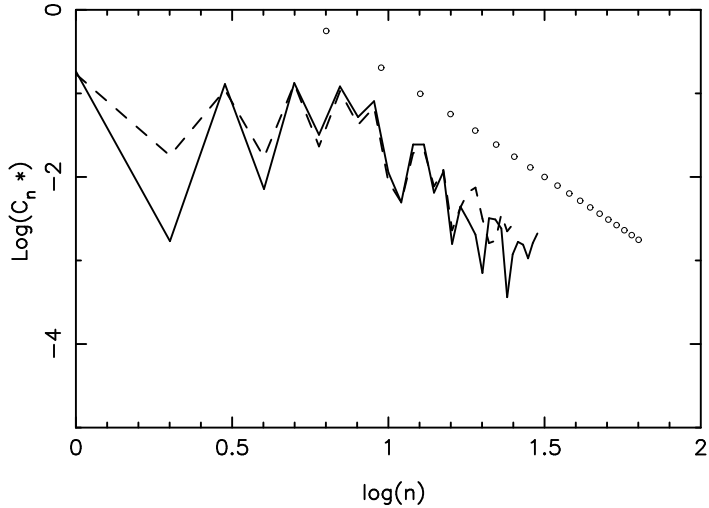}
\caption{{\small The Chebychev spectrum of the kinetic energy with $\epsilon =0$ at a time 40s after damping with initial particle spacing  1/125 (shown by the dashed line) and 1/150 (shown by the continuous line). The small circles mark the line $1/n^{2.5}$}.}
\end{center}
\end{figure}
 %------------------------------------------------------------------------

If this expansion is evaluated for the lines $x=\frac12$ and $y= \frac12$ and the results combined we can define a one dimensional spectrum 
\begin{equation}
E_K = \sum_n C^{\star}_n T^{\star}_n(x),
\end{equation}
where
\begin{equation}
 C^{\star}_n = \frac12 \sum_{i} (C_{ni} + C_{in} ) T^\star_i (1/2),
\end{equation}
where the summation has only even values of $i$. Some details of this calculation are given in the Appendix. 
      %----------------------------------------------------------------------
 \begin{figure}[h!]
 \begin{center}
\includegraphics[width=0.7 \textwidth ]{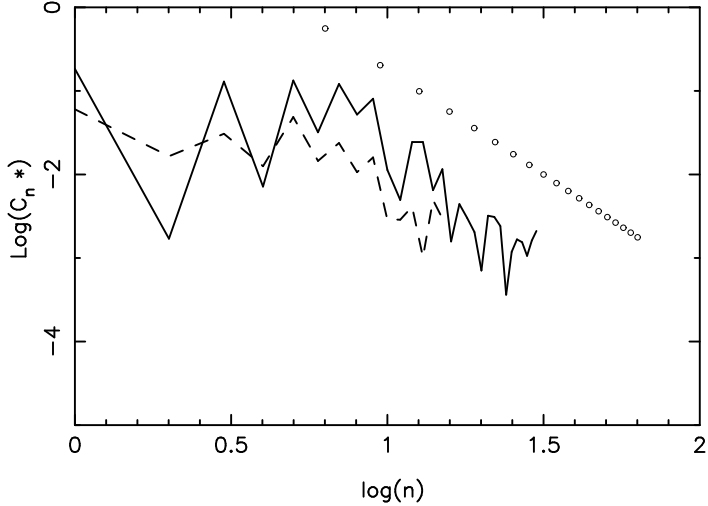}
\caption{{\small The Chebychev spectrum of the kinetic energy  for initial spacing 1/75 (with $\epsilon = 0.9$) shown by the dashed line), and 1/150 (with $\epsilon = 0.0$), shown by the continuous line). The small circles mark the line $1/n^{2.5}$}.}
\end{center}
\end{figure}
  
  %------------------------------------------------------------------------
    Figure 9 shows the Chebyshev spectrum when the resolution is 1/75 and $\epsilon$ has the values 0 (continuous line),  and 0.9 (dashed line).  As  in all the Chebyshev 1D spectra (including those of Clercx and Heijst 2000)  the spectrum has a pronounced zig-zag appearance because the modes with odd $n$ have coefficients that are larger than those for even $n$. The smoothing reduces the amplitude of the zig-zag.  The spectrum  is in satisfactory agreement with that of Clercx and van Heijst (2000) (see their figures 2(a)), though they calculate the spectra with higher resolution and a corresponding larger number of modes, and they average over two eddy turnover times.  Their $C^*_n$ fall off typically as $1/n^{p}$ where $2< p<3$ , and in figure 10 the line marked by small circles shows the case $p=2.5$.   The spectrum with $\epsilon = 0.9$ is significantly below that for $\epsilon = 0$ after mode number $\sim 10$. This mode number  is equivalent to a length scale of 0.1 or $5h $.

        Figure 10 shows the Chebyshev spectrum when the initial particle spacings are  1/150 and 1/125 and there is no smoothing.  This figure shows that the spectra are very similar for the two resolutions although the results for low mode numbers are more jagged for the 1/150 than for 1/125.  The agreement between these results confirms the inference from figures  5 and 7 that, at these resolutions, convergence is achieved.
    
    Figure 11   shows the spectrum with initial particle spacing 1/75 and $\epsilon = 0.9$ and the spectrum with initial particle spacing 1/150 and $\epsilon = 0$.   Although the spectrum for the spacing 1/150 runs to higher mode numbers than for 1/75, there is a strong suggestion that both spectra have a similar trend  for mode numbers $> 8$ but the averaged smoothed spectrum is lower, which is in agreement with the argument given in $\S 2$. No stronger conclusion can be drawn because the spectra are jagged.  It must be kept in mind that the one dimensional spectra depend on the where they are constructed. For example we could use lines parallel to the $y$ axis for $x$ equal to $\frac13$ and $\frac23$ with similar lines parallel to the $x$ axis. In this case the lines are closer to the no-slip boundary and the spectrum changes.  Clercx and Heijst (2000) consider the lines we use here as well as lines close to the boundaries where the spectrum falls off as $1/n^{5/3}$.        
%----------------------------------------------------------------------
\section{ Conclusions}
The results of this paper show  that SPH simulations using a standard algorithm reproduce the results obtained by other researchers for decaying turbulence in a square box with no-slip boundary conditions provided the resolution length is $<1/125$, in agreement with estimates made using the Reynolds number.  These results include the time variation of the decay of kinetic energy, enstrophy and the form of the spectra. The turbulence model proposed in this paper reproduces these results but, in addition, shows that quantities such as the second order velocity correlation function calculated using a resolution of 1/150, can be reproduced with an initial particle spacing 1/75 which requires a factor 8 less computation to integrate to a specified time.  

 The new algorithm is easy to implement and requires typically only $20 \%$ more computing time than a standard integration.  An attractive feature of the algorithm is that, like the LANS-$\alpha$ model,  it conserves linear and angular momentum (in the absence of fixed bodies or boundaries and external forces), and satisfies a discrete form of Kelvin's circulation theorem.  Detailed exploration of the  the dynamics for other boundary conditions, and other boundaries,  together with driven turbulence  is in progress.
 
%-----------------------------------------------------------------
\section { Acknowledgment}
This research was funded with the support of ARC Discovery grant DP0881447 (Analysis of two-phase sloshing in marine tanks).

\section{ References}
\begin{enumerate}
\item G.K.Batchelor, Computation of the Energy Spectrum in Homogeneous two-dimensional turbulence. Phys. Fluids.  Physica D,, Supp. II, 233-239, (1969).
\item S. Chen, S., Holm, D. D., Margolin, L. G., and Zhang, R. Direct numerical simulations of the Navier Stokes alpha model.Physica D. {\bf 133 } 66-83, (1999).

\item A. Cheskidov, Darryl D. Holm, Eric Olson and Edriss S. Titi. On a Leray-$\alpha$ model of turbulence. Proc. Phys. Soc. {\bf 461}, 629-649, (2005).

\item H.J.H. Clercx, S. R. Maassen, G.J.F. van Heijst. Decaying two-dimensional turbulence in square containers with no-slip or stree-free boundaries. Phys. of Fluids. {\bf 11(3)}, 611-626, (1999).

\item H. J. H. Clercx and G.J.F. Heijst. Energy spectra for decaying two dimensional turbulence in a bounded domain. Phys. Rev. Ltr. {\bf 85(2)}, 306-309, (2000).

\item S. Cummins and M. Rudman.  An SPH projection method. J. Computat. Phys. {\bf 152}, 584 607, (1999).

\item C. Eckart, C. Variation principles of hydrodynamics. Phys. Fluids, {\bf 3}, 421, (1960).

\item  Bernard. J. Geurts, and Darryl. D. Holm.  Leray and LANS-$\alpha$ modelling of turbulent mixing. J. Turb. {\bf 7(10)}, 1-33, (2006).

\item Jonathan Pietarila Graham, Darryl D. Holm, Pablo D. Miinni and Annick Pouquet, Highly turbulent solutions of the Lagrangian-averaged navier-Stokes $\alpha$ model and their large-eddy-simulation potential. Phys. Rev. E. {\bf 76}, 056310, (2007). 

\item Matthew Hecht, Darryl D. Holm, Mark R. Peterson, and Beth A. Wingate. Implementation of the LANS-$\alpha$ turbulence model in a primitive ocean model. J. Computat. Phys. $\bf 227(11)$, 5691-5716, (2008).

\item D. D. Holm, D. D. Fluctuation effects on 3D Lagrangian mean and Eulerian mean fluid motion.  Physica, {\bf 133}, 215 - 269, (1999).

\item D. D. Holm, Physica D, {\bf 170}, 253, (2002).

\item X.Y. Hu and N. A. Adams. An incompressible multi-phase SPH method. J. Computat. Phys. {\bf 227}, 264-278, (2007).

\item J. Kajtar, and J.J. Monaghan, SPH simulations of swimming linked bodies, J. Comput. Phys. 227 ,8568-8587, (2008).

\item E. Lunasin, S. Kurien, M.A. Taylor and E.S. Titi. A study of the Navier Stokes model for two dimensional turbulence. J. of Turbulence. {\bf 8(30)}, 1-21, (2007).

\item  Evelyn Lunasin, Susan Kurien, and Edriss S. Titi. Spectral scaling of the Leray-$\alpha$ model for two dimensional turbulence. J. Phys. A. Math. Theor. {\bf 41}, 344014, (2008).

\item S.R. Maassen, H.J.H. Clercx, and G.J.F. Heijst. Self organization of quasi-two dimensinal turbulence in stratified fluids in square and circular containers. Phys. of Fluids. {\bf14(7)}, 2150 - 2169. (2002).

\item K. Moheseni, B. Kozovic, S. Shkoller, and J. E. Marsden. Numerical simulations of the Lagrangian averaged Navier-Stokes equations for homogeneous isotropic turbulence.  Phys. Fluids. {\bf 15}, 524, (2003). 

\item J. J. Monaghan, On the problem of penetration in particle methods..  J. Computat. Phys., {\bf 82}, 1-15, (1989).

\item J. J. Monaghan. Smoothed Particle Hydrodynamics.   Ann. Rev. Astron. Ap., {\bf 30}, 543, (1992.)

\item J. J. Monaghan, Simulating free surface flows with SPH.  J. Computat. Phys., {\bf 110}, 399 -406, (1994).
\item J. J. Monaghan.  SPH Compressible turbulence. Mon.Not.Roy.Astro. Soc. {\bf 335}, 843-852, (2002).
\item J. J. Monaghan. Smoothed Particle Hydrodynamics. Rep. Prog. Phys. (2005).

\item J. J. Monaghan. SPH simulations of shear flow. Mon. Not. Roy. Astr. Soc. {\bf 365(1)}, 199-213, (2006).

\item C.S. Peskin, Numerical analysis of blood flow in the heart, J. Comput. Phys. {\bf 25}, 220-252, (1977).

\item C.S. Peskin, The immersed boundary method, Acta Numerica. {\bf 10 }, 479-517, (2002).

\item D. Violeau, and R. Issa.  Numerical modelling of complex turbulent free surface flows with the SPH method: an overview. Int. J. for Numerical Methods. {\bf 53}, 2077-, (2007).

\item H. Wendland. Piecewise polynomial, positive definite, and compactly supported radial functions of minimal degree.  Adv. Comput. Math. {\bf 4} 389-396, (1995).
\end{enumerate}

\section{Appendix}
 \setcounter{equation}{0}

\subsection{Boundary forces}
 In this paper we use the following form for the force per unit mass on fluid particle $c$ by boundary particle $j$, ${\bf f}_{cj}$, by 
 \begin{equation}
 {\bf f}_{cj} =  \frac{ 0.01c_s^2 B(r_{cj} )}{(|r_{cj}- d| ) } \frac{ {\bf r}_{cj} } { r_{cj} },
 \end{equation}
 where $B(r)$ is the one dimensional cubic Wendland function (Wendland  )which has been scaled so that it is only non zero in the domain $0<r/h<2$ where it has the form
 \begin{equation}
 \frac{1}{16}\left (2+3\frac{r }{h} \right ) \left (2- \frac{r}{h}\right )^3.
 \end{equation}
 The parameter $d$ is the spacing of the boundary force particles.  Provided $d$ is less than $0.5dp$, where $dp$ is the typical fluid particle spacing,  the tangential component of the force is $< 10^{-5}$ of the normal component.  Further details can be found in the paper by Monaghan and Kajtar (2009).
 
 \subsection{Chebyshev spectra}
 
 The continuum version of the SPH kinetic energy $E_K$ is
 \begin{equation}
\int _0^1\int _0^1E(x,y) dxdy =\frac 12 \int_0^1 \int _0^1 \rho {\bf v}({\bf r}) \cdot {\widehat {\bf v}}({\bf r}) dxdy
\end{equation}
The expansion of the kinetic energy per unit area $E(x,y)$ in Chebyshev polynomials $T_n^\star (q)$, such that $T_n^\star (q) = T_n (2q-1)$, with $0<q<1$, is given by
 \begin{equation}
  E(x,y) = \frac12\rho  {\bf v}(x,y) \cdot {\widehat {\bf v}}(x,y) =\sum_i \sum_j C_{ij} T_i^\star (x) T_j^\star (y),
\end{equation}
 where 
 \begin{equation}
N_{ij}C_{ij} =  \frac12 \int_0^1 \int_0^1 \frac{ \rho  {\bf v}(x,y)\cdot {\widehat {\bf v}}(x,y) T_i^\star (x) T_j^\star (y)dxdy}{\left(4x(1-x)y(1-y)\right)^{1/2}}.
\end{equation}
and
\begin{equation}
N_{ij} =  \int_0^1 \int_0^1 \frac{  T_i^\star (x)^2T_j^\star (y)^2dxdy}{\left(4x(1-x)y(1-y)\right)^{1/2}}.
\end{equation}
The denominator in these integrals is the weight function for the Chebyshev polynomials taking into account the change of argument given before (8.4). The first  integration can be approximated by an SPH summation and the second can be evaluated exactly. We get 

\begin{equation}
N_{ij}C_{ij} =  \frac12  \sum_a \frac {m_a {\bf v}_a \cdot {\widehat {\bf v}}_a T_i^\star (x_a) T_j^\star (y_a)}{\left(4x_a(1-x_a)y_a(1-y_a)\right)^{1/2}},
\end{equation}
as our estimate of the coefficients $C_{ij}$.  Tests on known functions shows that these estimates are sufficiently accurate for modes such that both $i$ and $j$ satisfy $i \le S/(5dp)$ where $dp$ is the intial particle spacing and $S$ is the length of a side of the square box.

\enddocument